LATeX, including 7 tables in laTeX.

\documentstyle[12pt]{article}
\begin{document}
\hsize 16cm
\baselineskip=20pt
\begin{center}
{\Large \bf 5D Axisymmetric Stationary\\ Solutions as Harmonic Maps}
\end{center}
\begin{center}
by\\
Tonatiuh Matos$^{+}$ \\
Departamento de F\'{i}sica \\
Centro de Investigaci\'{o}n y de Estudios Avanzados\\
del I.P.N.\\
A.P. 14-740, C.P. 07000\\
M\'{e}xico, D.F.\\
\end{center}
\vskip 2cm
Abstract.  We present the complete scheme of the application of the
one-and two dimensional subspace and subgroups method to five-dimensional
gravity with a $G_{3}$ group of motion.  We do so in the space time and
in the potential space formalisms.  From this method one obtains the Kramer,
Belinsky-Ruffini, Dobiasch-Maison, Cl\'{e}ment, Gross-Perry-Sorkin solutions
etc. as special cases.
\vskip 2cm
PACS: 04.50.+th, 04.20.Jb.
\vskip 3cm
+Work supported in part by CONACYT-M\'{e}xico.
\vfil\eject
\section{Introduction.}
\indent
High-dimensional relativity is perhaps one of the oldest and most elegant
ways for unifying all interactions in physics.  The first idea has been more
and more transformed from the original suggestion of Kaluza and Klein$^{1}.$
In this work we are interested in the 5D theory (5-dimensional relativity),
which unifies gravitation with electromagnetism.  We adopt the version where
the
5D Riemannian space P is a principal fibre bundle$^{2}$ with typical fibre
S$^{1},$ the circle.  This version is more convinient because 1) it is a
natural generalization of U(1)-gauge theory (Maxwell-theory) to curved
spacetimes$^{2},$ 2) we do not need to recur to the so-called Kaluza-Klein
ansatz$^{3}$ 3) we do not need to impose any restrictions to the functional
dependence of the metric terms on P and 4) we do not need to use the so-called
n-mode ansatz$^{3}.$ Let us shortly explain these affirmations.

Let $U\subset B^{4}$ be an open subset of $B^{4}=P/S^1$ and $\pi$ the
projection
$\pi :P\rightharpoonup B^{4}. B^{4}$ is interpreted as the spacetime
with metric$^{4}g.$ If $\phi : P\rightharpoonup U\times S^{1}$ is a
trivialization of an open subset of P then we can map the physical quantities
into the set $U\times S^{1}$ through the trivialization $\phi .$  Let
${\tilde g} = I^{2}{\hat\omega}\otimes {\hat\omega}$ be a metric on
$S^{1}$ and ${\hat g} = \eta_{AB}{\hat\omega}^{A}\otimes {\hat\omega}^{B} =
\eta_{\mu\nu}{\hat\omega}^{\mu}\otimes{\hat\omega}^{\nu} +I^{2}
{\hat\omega}\otimes{\hat\omega} \,\,A,B =1,...,5 \ ; \mu ,\nu = 1,...,4$ be the
metric on P = H$\oplus$ V, where V is the vertical space of P and H its
complement.  (To choose H is equivalent to choose the Maxwell connection).
Remmember that if $\{{\hat\omega}^{A}\}\subset T^{*}P$ is a base of the
cotangent space of P with dual $\{{\hat e}_{A}\}\subset TP$ then the vertical
space is defined as the set ${\hat e}\epsilon TV$ such that
$d\pi ({\hat e} ) =0.$  In this case of course TV is one-dimensional.  Let
$\{e_{\alpha}\}$ be the projection of the complement base
$\{ {\hat e}_{\alpha}\}$, i.e $d\pi ({\hat e}_{\alpha} ) = e_{\alpha}$ and
$\pi_{1}$ be the first projection, $\pi_{1} : U\times S^{1} \rightharpoonup
 U, \pi_{1}(x,\theta ) =x.$  Thus, because of the identity $\pi =
\pi_{1} o \phi$ , one finds that the corresponding base in the trivialization
set $U\times S^{1}$ is $d\phi (\{{\hat e}_{\alpha},{\hat e}\})=
\{e_{\alpha} -A_{\alpha}{\partial\over{\partial\theta}},
{\partial\over{\partial\theta}}\}$ whose dual base is $\{\omega^{\alpha} ,
d\theta + A_{\alpha}\omega^{\alpha}\}^{5}$ .  We can now write down the
corresponding metric $\bar g$ on $U \times S^{1}$ :
\begin{equation}
{\bar g}=\eta_{\mu\nu}\omega^{\mu}\otimes\omega^{\nu} + I^{2}
(A_{\alpha}\omega^{\alpha}+d\theta )\otimes(A_{\beta}\omega^{\beta} +
d\theta )
\end{equation}
whose pullback into P is just $\phi^{*}{\bar g}={\hat g}.$  Let ${\hat A}$
be the one-form of connection on P, (it is easy to see that ${\hat A} =
{\hat\omega})$ and $s$ be a cross section $s:U \rightharpoonup P.$  If
${\bar Id}:U\rightharpoonup U\times S^{1};x\rightharpoonup (x,0),s$
can allways be written as $s=\phi^{-1}\circ{\bar Id}.$  With this last
relation it can be shown that $s^{*}({\hat\omega} )=A_{\alpha}\omega^{\alpha}.$
That means that $A_{\alpha}$ are the pullback components of the one-form of
connection ${\hat A}$ through a cross section $s$.  Observe that metric (1) is
not as ansatz.  Since the group $U(1)\cong S^{1}$ is acting on P
there exist an isometry Is: $P\rightharpoonup P, (x^{\mu},\theta )
\rightharpoonup Is(x^{\mu},\theta )=(x^{\mu},\theta +2\pi)$ such that
$Is^{*}{\bar g}={\bar g}.$  This implies the existence of a Killing-vector
$X$ , therefore we can work on the coordinate system where
$X={\partial\over{\partial\theta}}.$
In this coordinate system the metric components of ${\bar g}$ do not
depend on $x^{5}=\theta.$  Observe that the no $x^{5}$ dependence of the
${\bar g}$ components is consequence of the action of $U(1)$ on P.
With the gauge-theory philosophy the action of $U(1)$ on P means that
there are electromagnetic interactions on $P^{6}.$  Thus due to this
philosophy the existence of electromagnetic interactions implies that there
is a coordiante system (a chart on P) in which the metric components do not
depend on $x^{5}$, (the n=0 mode ansatz is not an approximation but the choise
of a chart$^{5,6}$ ).
In this work we deal with 5D gravity, ${\hat R}_{AB}=0$ , where the
metric ${\bar g}$ depends only on two coordinates, i.e. a $G_{3}$ group
motion is acting on P. (Solutions for higher-dimensional spaces P, can be
found for example in ref.[45]).
  This can be done in two formalism 1) in the spacetime
and 2) in the potential space formalism (analogous to the Ernst
potentials$^{7}).$  The main goal of this work is to develop a method for
generating exact solutions of the 5D Einstein field equations in vacuum
when the metric ${\bar g}$ depends on two coordinates.  We will reduce
these exact solutions to harmonic maps.  First we are going to work in the
spacetime.
\section{Spacetime Field Equations}
\indent
In this section we start with the metric (1) when ${\bar g}$ depends on two
coordinates $\rho$ and $\zeta$ . In this case we can write $\bar g$ as
\[
{\bar g}=H(d\rho^{2} + d\zeta^{2} )+{\bar g}_{ij}dx^{i}dx^{j}
\]
\begin{equation}
i,j =3,4,5
\end{equation}
where ${\bar g}_{ij}={\bar g}_{ij}(\rho ,\zeta )$ and $H=H(\rho, \zeta ).$
Now we define a 3x3-matrix $\gamma$ whose components are just the
${\bar g}_{ij}$ metric components, $(\gamma )_{ij}={\bar g}_{ij}.$ In terms
of the metric $g$ of $B^{4}$ and the connection components
$A_{\mu},\gamma$ reads
\begin{equation}
\gamma =\left(\begin{array}{ccc}
         {1\over{I}}g_{33}+I^{2}A^{2}_{3}&
         {1\over{I}}g_{34}+I^{2}A_{3}A_{4}&
         I^{2}A_{3}\\
         {1\over{I}}g_{34}+I^{2}A_{3}A_{4}&
         {1\over{I}}g_{44}+I^{2}A^{2}_{4}&
         I^{2}A_{4}\\
         I^{2}A_{3}&I^{2}A_{4}&I^{2}
\end{array}\right)
\end{equation}
where we have normalized $g_{\mu\nu}\rightarrow {1\over{I}}
g_{\mu\nu}$ in order to have det$\gamma =(g_{33}g_{44}-g_{34}^2) =-
\alpha^{2}.$ It is also argued that the normalized $g_{\mu\nu}$ and not the
original ones are the physical gravitational potentials$^{6,8}.$  In terms of
this matrix the Einstein field equations in vacuum ${\bar R}_{AB}=0,\, A,B=
1,...,5$ are $^{9,10}$

\noindent
a)\[({\it ln}H)_{,z}={({\it ln}{\alpha})_{,zz}\over
                       {({\it ln}{\alpha})_{,z}}}+
                       {1\over{4({\it ln}{\alpha})_{,z}}}
tr(\gamma_{,z}\gamma^{-1})^{2}\]
b)\[
               (\alpha \gamma_{,z}\gamma^{-1})_{,{\bar z}}+
               (\alpha \gamma_{,{\bar z}}\gamma^{-1})_{,z}=0 \]
\begin{equation} z=\rho + i\zeta
\end{equation}
Equation (4.b) implies
\begin{equation}
 \alpha_{,z{\bar z}}=0
\end{equation}
The solution of equations (4) will depend on the solution of equation (5).
The choice $\alpha ={1\over{2}}(z+{\bar z})=\rho$ corresponds to the Weyl's
canonical coordinates.
To solve equation (4.a) we need first a solution of (4.b).  Let us define
the 3x3-matrix g $=-\alpha^{-2/3}\gamma$ . g is a matrix
of the group $SL(3,{\cal R})$ which fulfills
\begin{equation}
(\alpha\hbox{g}_{,z}\hbox{g}^{-1})_{,{\bar z}}+
(\alpha\hbox{g}_{,{\bar z}}\hbox{g}^{-1})_{,z}=0
\end{equation}
Then it is of great interest to solve the chiral equation (6), which will
solve equations (4).
\section{Potential space field equations.}
\indent
The potential formalism was first introduced by Neugebauer$^{11}$ and
rediscovered by Maison$^{12}$ ten years later.  It consists in defining
covariantly five potentials in terms of the two commutating Killing vectors
X and Y, X being related to the U(1) isometry and Y being related to
stationarity.  The five potentials are$^{11}$
\[
I^{2}=\kappa^{4/3}=X_{A}X^{A} \;\;\;\;\;\; f=-IY^{A}Y_{A}+
                                           I^{-1}(X^{A}Y_{A})^{2} \;\;\;\;\;
\psi =-I^{-2}X_{A}Y^{A}
\]
\begin{equation}
\epsilon_{,A}=\epsilon_{ABCDE}X^{B}Y^{C}X^{D;E} \;\;\;\;
\chi_{,A}=\epsilon_{ABCDE}X^{B}Y^{C}Y^{D;E}
\end{equation}
being $\epsilon_{ABCDE}$ the Levi-Civita pseudotensor. In the adapted
coordinate
system where $X=X^{A}{\partial\over{\partial x^{A}}}=
{\partial\over{\partial x^{5}}},\, Y=Y^{A}{\partial\over{\partial x^{A}}}=
{\partial\over{\partial x^{4}}}$ , one finds that $f, \epsilon,\psi ,\chi ,
\kappa$ are the gravitational, rotational, electrostatic, magnetostatic and
scalar potentials, respectively.  The field equations (4.b) in terms of the
five potentials $\Psi^{A}=(f,\epsilon ,\psi ,\chi ,\kappa )$ are five
non-linear, second-order, partial differential equations which can be derived
from the Lagrangian$^{11,13}$
\[
{\cal L}={\alpha\over{2f^{2}}}[f_{,i}f^{,i}+(\epsilon_{,i}+\psi\chi_{,i})
(\epsilon^{,i}+\psi\chi^{,i})]+
\]
\begin{equation}
{\alpha\over{2f}}(\kappa^{2}\psi_{,i}\psi^{,i}+{1\over{\kappa^{2}}}
\chi_{,i}\chi^{,i})+{2\over{3}}{\alpha\over{\kappa^{2}}}
\kappa_{,i}\kappa^{,i}
\end{equation}
(variantion with respect to $\Psi^{A}$ ). Now we can define a five-dimensional
(abstract) Riemannian space $V_{5}$ inspired on the Lagrangian (8) with
metric
\begin{equation}
dS^{2}={1\over{2f^{2}}}[df^{2}+(d\epsilon +\psi d\chi )^{2}]+
{1\over{2f}}(\kappa^{2}d\psi^{2}+{1\over{\kappa^{2}}}d\chi^{2})+
{2\over{3}}{d\kappa^{2}\over{\kappa^{2}}}.
\end{equation}
On $V_{5}$, the five potentials $\Psi^{A}$ are the local coordiantes
which define a symmetric Riemannian space (all covariant derivatives of the
Riemann tensor vanish).
\indent
We call $V_{5}$ the potential space.  In the axisymmetric case there exist
a third Killing vector Z forming a $G_{3}$ group of motion on P.  Like
in the last section, we use the complex coordinate $z=\rho +i\zeta$ .  The
field equations derived from (8) can be cast into first order differential
equations.  In order to do so we define$^{13}$
\[
A_{1}={1\over{2f}}[f_{,z}-i(\epsilon_{,z}+\psi \;\; \chi_{,z})] \;\;\;\;\;
D_{1}={1\over{3}}{\kappa_{,z}\over{\kappa}}
\]
\[
B_{1}={1\over{2f}}[f_{,z}+i(\epsilon_{,z}+\psi \;\; \chi_{,z})] \;\;\;\;\;
C_{1}=({\it ln}\kappa )_{,z}
\]
\[
E_{1}=-{1\over{2\sqrt{2}}}f^{-1/2}[\kappa \; \psi_{,z}-i{\chi_{,z}\over
{\kappa}}]
\]
\begin{equation}
F_{1}={1\over{2\sqrt{2}}}f^{-1/2}[\kappa \; \psi_{,z}+{i\chi_{,z}\over
{\kappa}}]
\end{equation}
and $A_{2},B_{2}...$ etc. with ${\bar z}$ in place of $z$ .  In terms of
these quantities the field equations read$^{13}$
\[
A_{1,{\bar z}}=A_{1}A_{2}-A_{1}B_{2}-{1\over{2}}C_{2}A_{1}-{1\over{2}}
C_{1}A_{2}-2E_{1}F_{2}
\]
\[
A_{2,z}=A_{1}A_{2}-A_{2}B_{1}-{1\over{2}}C_{2}A_{1}-{1\over{2}}C_{1}
A_{2}-2E_{2}F_{1}
\]
\[
B_{1,{\bar z}}=B_{1}B_{2}-A_{2}B_{1}-{1\over{2}}C_{2}B_{1}-{1\over2}C_1B_2-
2E_{2}F_{1}
\]
\[
B_{2,z}=B_{1}B_{2}-A_{1}B_{2}-{1\over{2}}C_{2}B_{1}-{1\over{2}}
C_{1}B_{2}-2E_{1}F_{2}
\]
\[
E_{1,{\bar z}}=A_{1}E_{2}+{1\over{2}}A_{2}E_{1}-{1\over{2}}
B_{2}E_{1}-{1\over{2}}C_{1}E_{2}-{1\over{2}}C_{2}E_{1}+3D_{1}F_{2}
\]
\[
E_{2,z}=A_{2}E_{1}+{1\over{2}}A_{1}E_{2}-{1\over{2}}B_{1}E_{2}
-{1\over2}C_1E_2-{1\over{2}}C_{2}E_{1}+3D_{2}F_{1}
\]
\[
F_{1,{\bar z}}=B_{1}F_{2}+{1\over{2}}B_{2}F_{1}-{1\over{2}}
A_{2}F_{1}-{1\over{2}}C_{1}F_{2}-{1\over{2}}C_{2}F_{1}+3D_{1}E_{2}
\]
\[F_{2,z}=B_{2}F_{1}+{1\over{2}}B_{1}F_{2}-{1\over{2}}A_{1}F_{2}-
{1\over2}C_1F_2-{1\over{2}}C_{2}F_{1}+3D_{2}E_{1}
\]
\[D_{1,{\bar z}}=-(E_{1}E_{2}+F_{1}F_{2})-{1\over{2}}C_{1}D_{2}-
{1\over{2}}C_{2}D_{1}
\]
\begin{equation}
D_{2,z}=-(E_{1}E_{2}+F_{1}F_{2})-{1\over{2}}C_{1}D_{2}-{1\over{2}}
C_{2}D_{1}
\end{equation}
We have now ten non linear first order differential equations in place
of five of second order.  The important point is that we can find an
equivalent linear problem (Lax pairs-representation) for equations (11),i.e.
for the axisymmetric stationary field equations (4) in the potential space.
It reads$^{13}$
\[
\Omega_{,z}=[\alpha_{1}+\lambda\beta_{1}]\Omega
\]
\[
\Omega_{,{\bar z}}=[\alpha_{2}+{1\over{\lambda}}\beta_{2}]\Omega
\]
\[
\lambda =\sqrt{\frac{K-i{\bar z}}{K+iz}}
\]
\begin{equation}
\alpha_i=\left(
\begin{array}{lll}
B_{i}&0&E_{i}\\
0&A_{i}&-F_{i}\\
-F_{i}&E_{i}&\frac{1}{2}(A_{i}+B_{i})
\end{array}
\right),\;\;\; \beta_{i}  =
\left(
\begin{array}{rrr}
-D_{i}&B_{i}&F_{i}\\
A_{i}&-D_{i}&-E_{i}\\
E_{i}&-F_{i}&2D_{i}
\end{array} \right)_{i=1,2}
\end{equation}
Equations (11) are equivalent to the integrability conditions of the 3x3
matrix $\Omega$ .  The solitonic method (generalized inverse scattering
method) was applied to the linear problem (12) for finding B$\ddot a$cklund
transformations and generating new solutions from seed ones$^{14}$ .
Linear problem (12) is invariant under an 8-parametric group of
transformations:
This group is $SL(3,{\bf R})^{12,15}$ , which is the group of isometries
of the metric (9).  Then it is possible to write metric (9) as
$dS^{2}={1\over{4}}tr(d\hbox{g}d\hbox{g}^{-1})$ where $\hbox{g}\epsilon
SL(3,{\bf R}).$  A parametrization of g in terms of the potentials
$\Psi^{A}$ was first given by Maison$^{12}$ and nine years later by the
author$^{15}.$ g can be parametrized as
\begin{equation}
\hbox{g}=\frac{-2}{f\kappa^{\frac{2}{3}}}\left(
\begin{array}{lll}
f^2+\epsilon^2 -f\kappa^2\psi^2&-\epsilon &-\frac{1}{2\sqrt{2}}(\epsilon
\chi+f\kappa^2\psi) \\
-\epsilon&1& \frac{1}{2\sqrt{2}}\chi \\
-\frac{1}{2\sqrt{2}}(\epsilon\chi+f\kappa^2\psi)&\frac{1}{2\sqrt{2}}\chi&
\frac{1}{8}(\chi^2-\kappa^2f)
\end{array}\right)
\end{equation}
with g, the field equations (11) are just the chiral equations (6)
which are invariant under $SL(3,{\bf R})$ -transformations
\begin{equation}
\hbox{g}\rightharpoonup\hbox{g}^{'}=c\hbox{g}c^{T}
\end{equation}
being $c\epsilon SL(3,{\bf R})_{c},i.e.$ $c$ is a constant matrix of
$SL(3,{\bf R}).$  So we have to solve the chiral equations (6) in the
potential space, too.  In the next section we give a method for solving these
equations.
\section{ $SL(3,{\bf R})$ -invariant chiral equations. }
\indent
Chiral equations has been material of great study in the last years$^{16}$ .
B$\ddot a$cklund transformations have been applied for finding explicit
chiral fields of an arbitrary group $G^{17}$ .  In this work we will use an
alaternative method developed in a previous paper$^{18}.$  It has been used
in four dimensions, where we find a natural classification of the solutions
in classes $^{19}.$  It reduces the chiral equations to harmonic maps.  We
shall outline it in this section.
Let $\lambda^{i} \;\; , i=1,...,p $ be harmonic maps$^{20}$
\begin{equation}
(\alpha\lambda^{i}_{,z})_{,{\bar z}}+
(\alpha\lambda^{i}_{,{\bar z}})_{,z}+2\alpha{\Gamma^{i}_{jk}}
\lambda^j_{,z}{\lambda^{k}_{,{\bar z}}}=0
\end{equation}
where $\Gamma^{i}_{jk}$ are the Christoffel symbols of a Riemannian space
$V_{p}.$  Suppose that $\hbox{g}=\hbox{g}(\lambda^{i})\subset G$. This ansatz
was first used by Neugebauer and Kramer in the
Einstein-Maxwell theory$^{21}.$  If we do so, the chiral equations (6)
reduce to the Killing equations on $V_{p}$ for the components of the
Maurer-Cartan forms $A_{i}=(\partial_{\lambda^{i}}\hbox{g})\hbox{g}^{-1}$
\begin{equation}
A_{i;j}+A_{j;i}=0
\end{equation}
with
\begin{equation}
A_{i;j}={1\over{2}}[A_{j},A_{i}]
\end{equation}
where [ ] means matrix commutator.  If g$\epsilon G$ is a Lie group
matrix, then $A_{i}\epsilon {\cal G}\;\;({\cal G}$
is the corresponding Lie algebra of $G$).
Let $\{\xi_{j}\}$ be a base of the Killing vectors on $V_{p}$ , and
$\{\sigma_{j}\}$ a base of the Lie algebra ${\cal G}$ .  Thus we can write
$A_{i}$ as
\begin{equation}
A_{i}=\xi^{j}_{i}\sigma_{j}
\end{equation}
\indent
The relation: $A^{c}_{i}\sim A_{i}$ iff there exist $c\epsilon G_{c}$(the
group of constant matrices in G) such that $A^{c}_{i}=A_{i}oL_{c},$ where
$L_{c}$ is the left action of $G_{c}$ over G, is an equivalence relation
which separates the set of matrices of ${\cal G}$ into equivalence classes,
$\{A_{i}\}/\sim =\{[A_{i}]\}.$  Let TB be a set of representatives of
$\{[A_{i}]\}.$  Of course TB$\subset{\cal G}$ .
We can map the elements of TB into $G$,
through the exponential map or by integration.  Define
$B=\{\hbox{g}\epsilon G\vert\hbox{g}= exp A_{i}, \;\;for\;\;all\;\;
A_{i}\epsilon TB\}\subset G.$ Then the following theorem holds.  The sextet
$(G, B, \pi, G_{c}, U, L_{c})$ where $U$ is a local coordinate system of
G, and $\pi$ is the projection $\pi(L_{c}(\hbox{g}))=\hbox{g},$ constitutes
a principal fibre bundle (for the details of the demostration see ref. [18]).
That means that it is enough to obtain the base set B for generating all the
set of solutions by left action $L_{c}.$  We are interested in the group
$G=SL(3,{\bf R}),$ eather for the space time as for the potential space.
The base space B for the one-and two- dimensional subspaces i.e. for
p=1 and p=2 in (15) is given in ref. [22] and [23].  In table I we reproduce
the corresponding set B for the one-parametric subgroups$^{22}$ and in table
II for the abelian two-parametric subgroups of $SL(3,{\bf R})$ where the
parameters $\lambda$ and $\tau$ fulfill the Laplace equation
\begin{equation}
(\alpha\tau_{,z})_{,{\bar z}}+(\alpha\tau_{,{\bar z}})_{,z}=0 \;\; ; \;\;
(\alpha\lambda_{,z})_{,{\bar z}}+(\alpha\lambda_{,{\bar z}})_{,z}=0
\end{equation}
(for details and calculations see ref. [23]).  The classical two-dimensional
subgroups of $SL(3,{\bf R})$ are $SL(2,{\bf R})$, $Sp(2,{\bf R})$ and
$SO(2,1). SL(2,{\bf R})$ and $Sp(2,{\bf R})$ have the same algebra
representation and therefore we have the same solution
\begin{equation}
\hbox{g} =  \frac{1}{1-\lambda\tau}\left(
\begin{array}{lll}
c(1-\lambda )(1-\tau) &e(\tau - \lambda )&0\\
e(\tau - \lambda )&d((1+\lambda)(1+\tau)&0\\
0&0&1-\lambda\tau
\end{array}\right)
\end{equation}
while for the group $SO(2,1)$ we arrive at
\begin{equation}
\hbox{g}=  \frac{1}{(1-\lambda\tau)^2}
\left(
\begin{array}{lll}
a(\lambda-1)^2(\tau-1)^2&
b(\lambda-\tau)^2&c(\lambda-1)(\tau-1)(\lambda-\tau)\\
b(\lambda-\tau)^2 &d(\tau +1)^2(\lambda
+1)^2&e(\lambda+1)(\tau+1)(\lambda-\tau)
\\
c(\lambda -1)(\tau -1)(\lambda -\tau)&e(\lambda+1)(\tau+1)(\lambda-\tau)&
f[(1-\lambda\tau)^2 -2(\tau -\lambda)^2]
\end{array} \right)
\end{equation}
(we have transformed $\lambda\rightharpoonup 2\lambda$ and $\tau\rightharpoonup
2\tau$ from ref. [23]).  Here the parameters $\lambda$ and $\tau$ build minimal
surfaces on P (see eq. (15))
\[
(\alpha\lambda_{,z})_{,{\bar z}}+(\alpha\lambda_{,{\bar z}})_{,z}+
{4\alpha\tau\over{1-\lambda\tau}}\lambda_{,z}\lambda_{,{\bar z}}=0
\]
\begin{equation}
(\alpha\tau_{,z})_{,{\bar z}}+(\alpha\tau_{,{\bar z}})_{,z}+
\frac{4\alpha\lambda}{1-\lambda\tau}\tau_{,z}\tau_{,{\bar z}}=0
\end{equation}
Solutions of $\lambda$ and $\tau$ in (22) in terms of $z$ and $\bar z$ are
well-known.  A method for integrating equations (22) is given in [24].  So
for each solution of (19) respectively of (22) we have a new solution of (6).
Furthermore from each solution of tables I and II or (20) and (21) we can
obtain a new solution by left action of the group $SL(3,{\bf R})_{c}.$  In
our case, matrix g is symmetric in both representations, spacetime
and potential space.  Therefore the left action
\begin{equation}
L_{c}(\hbox{g})=c\hbox{g}c^{T}
\end{equation}
is more convinient in order to get $L_{c}(\hbox{g})$ also symmetric.
To end this section it is easy to show that there exist a superpotential
$\beta$ defined as$^{25}$
\begin{equation}
\beta_{,z}={1\over{4({\it ln}\alpha )_{,z}}}tr(\hbox{g}_{,z}\hbox{g}^{-1})^{2}.
\end{equation}
Indeed, chiral equations are just the integrability conditions of $\beta$ .
This superpotential $\beta$ will play an important role in the next section.
\section{The unifying point of view}
\indent
It is clear that there exist a direct relation between solutions in
spacetime and in the potential space, i.e. if we have an exact solution in the
spacetime we can get a different one in potential space and viceversa using the
identity
\[
\frac{1}{\alpha^{\frac{2}{3}}}
\left( \begin{array}{lll}
\frac{1}{I}g_{33}+I^2A^2_3 & \frac{1}{I}g_{34}+I^2 A_3 A_4 &I^2 A_3\\
\frac{1}{I}g_{34}+I^2A_3A_4 & \frac{1}{I}g_{44}+I^2 A^2_4  &I^2 A_4\\
 I^2 A_3 & I^2 A_4  & I^2
\end{array}\right) =
\]
\begin{equation}
\frac{2}{f\kappa^{\frac{2}{3}}}
\left(
\begin{array}{lll}
f^2+\epsilon^2 -f\kappa^2\psi^2& -\epsilon& -\frac{1}{2\sqrt{2}}\left(\epsilon
\chi+f\kappa^2\psi \right) \\
-\epsilon&1& \frac{1}{2\sqrt{2}}\chi\\
-\frac{1}{2\sqrt{2}}\left(\epsilon\chi+f\kappa^2\psi\right)&
\frac{1}{2\sqrt{2}}\chi& \frac{1}{8}\left(\chi^2-\kappa^2f\right)
\end{array} \right) = \hbox{g}
\end{equation}
\indent
There exist an analogous relation in four-dimensional gravity even through it
is no so direct, it is complex. (The relation between quantities in
four-dimensional gravity in spacetime and potential space was
first given in ref.[21] and used later in ref. [26].  See also ref. [19]).
\indent
The next step is comparing matrices (25) with matrices g in tables
I and II, (20) and (21) to write down the potentials and the elements of the
metric tensor in terms of the harmonic maps $\lambda$ and $\tau$ .
Some results are even given in the literature. Here we complete them.
The one-dimensional subspaces in spacetime are given in Table III.
All of them represent new solutions.
One-dimensional subspaces in potential-space was first given in ref. [27] for
matrices $A_{i}$ fulfilling $trA^{2}_{i}=0$, i.e: $\beta =0$ in table I and II.
The complete set of solutions of this subspace in the potential space is
given in ref. [28] and we reproduce the results in table IV.
\indent
Some abelian two-dimensional subspaces in the potential space are also
given in [27] (for $\beta =0$ ). The whole set of abelian two-dimensional
subspaces in the space time are given in table V.
All of them are again new solutions of the field equations.
Here we complete the set of solutions from the two-dimensional subspaces
in the potential space and write all of them
in table VI.  Finally we write the results for both representations for the
groups $SL(2,{\bf R})\cong Sp(2,{\bf R} )$ and $SO(2,1)$ in table VII.
All the spacetime solutions in this table are new.
With this we solve equations (4.b).  To solve equation (4.a) in the spacetime
we substitute the relation $\gamma =-\alpha^{2/3}\hbox{g}$ in it to obtain.
\begin{equation}
\left( ln\frac{H\alpha^{\frac{2}{3}}}{\alpha_{,z}}\right)_{,z}=
\frac{1}{2\alpha_{,z}}\frac{1}{2}\alpha tr(\hbox{g}_{,z}\hbox{g}^{-1})^2
\end{equation}
Of course we need the expresion (26) in terms of the harmonic maps at the
right hand side, i.e.
$\hbox{g}_{,z}\hbox{g}^{-1}=\hbox{g}_{,\lambda^i}
\hbox{g}^{-1}\lambda^{i}_{.z}=A_{i}\lambda^{i}_{,z}.$

 If we do so,
for the one-dimensional subspaces equation (26) reduces to
\begin{equation}
({\it ln}H\rho^{2/3})_{,z}={1\over{2}}\rho trA^{2}(\lambda_{,z})^{2}
\end{equation}
where we have used Weyl's coordinates $\alpha ={1\over{2}}(z+{\bar z})=\rho$
for convenience, while for the two-dimensional abelian subspaces (26) reduces
to
\begin{equation}
({\it ln}H\rho^{2/3})_{,z}={\rho\over{2}}[tr\sigma^{2}_{1}(\lambda_{,z})^{2}
+tr\sigma^{2}_{3}(\tau_{,z})^{2}+2 \; tr(\sigma_{1}\sigma_{3})
\lambda_{,z}\tau_{,z}]
\end{equation}
For the $SL(2,{\bf R})$ subspaces equation (26) reads$^{19}$
\begin{equation}
({\it ln}H\rho^{2/3})_{,z}= -2\rho\frac{(\lambda -\tau)^{2}}{
(1-\lambda\tau )^{4}}\lambda_{,z}\tau_{,z}
\end{equation}
Finally for the group SO(2,1) equation (26) reads
\begin{equation}
({\it ln}H\rho^{2/3})_{,z}={4\rho\over{(1-\lambda\tau )^{2}}}
\lambda_{,z}\tau_{,z}.
\end{equation}
\indent
In the space time the left action of $SL(3,{\bf R})_{c}$ corresponds to gauge
transformations of the metric, but in the potential space it generates really
new solutions $^{11,29}.$  The left action of the group in the potential
space are actually Ehlers-Harrison type transformations for 5D
gravity $^{12,15}$ .  In the next section we will give some examples of such
transformations.
\section{Examples}
\indent
In this section we want to recover some of the well-known exact solution from
tables III-VII.

\indent
a) Heckmann - Jordan - Fricke and K$\ddot u$hnel-Schmutzer solutions.

The cases a), b) and $q \not= 0$ of table IV was studied in ref. [30] where
monopoles, dipoles, quadripoles solutions was derived.  We give other
examples.  We take the case a) of table IV with $r_{1}=1/H, r_{2}=-1,
r_{3}=1-1/H, I_{0}=2ac$ and $\lambda = {\it ln} V.$  We obtain the
metric$^{28}$
\[
dS^{2}=e^{\sigma}dr^{2}+r^{2}(d\theta^{2}+sin^{2}\theta d\phi^{2})-
e^{\nu}dt^{2}+I^2dx^{2}
\]
\[
r=r_{0}[e^{{1\over2}\lambda}(e^{-h\lambda}+e^{h\lambda})]^{-1} \;\;\;\;\;
h^{2}H^{2}={1\over{4}}(H^{2}-H+1)
\]
\begin{equation}
e^{\nu}=V^{1/H} \;\;\; , e^{\sigma}=4h^{2}[(h+{1\over{2}})
e^{h\lambda}+(h-{1\over{2}})e^{-h\lambda}]^{-2}; \;\;\;
I=I_{0}V^{{1\over2}(1-{1\over{H}})}
\end{equation}
This metric is known as the Heckmann - Jordan - Fricke solution$^{31}$ .  If
we make a (23)-like transformation where the matrix A in case a) of Table IV
transforms to$^{28}$
\[
cAc^{-1}=A^\prime=\left(
\begin{array}{lll}
h_1&0&h_2\\
0& -(h_1+h_4)&0\\
h_3&0&h_4
\end{array}
\right)
\]
and define
\[
h_{2}=\sqrt{2}(h_1-h_4)[ hd P_{1}+\frac{ec}{I^2_0}P_{2}]
\]
\[
h_{3}=\frac{4(h_1 - h_4^2)cdeh}{h_2(I^2_0+16cdeh)}
\]
$$
P_{1,2}=1\pm
\sqrt{ \frac{I^0_2}{I^2_0+16cdeh}} \;\;\;\;\;\;\;\;
I_0, c, d, e, h, = const
$$
one obtains the K$\ddot{u}$hnel-Schmutzer solution $^{32}$ (see also ref.
[33])

\indent
b) Kramer - and Belinsky - Ruffini solutions.

\indent
Einstein theory in vacuum belongs to the $SL(2,{\bf R})$
subspaces.  For example, if we choose
$\xi =\frac{m+i{\it l}}{r-m+ia \;\; cos\theta}$
in table VII in the potential
space we obtain the
Kerr-NUT solution.  Furthermore in equations (11) we can see that if we set
$\psi =\chi =0$ , the $\kappa$ equation becomes uncoupled from the others.  Let
us start from the Kerr-NUT solution and a function $\kappa$ given by$^{15}$
\[
\kappa =\left(\frac{r-m+\sigma}{r-m-\sigma}\right)^{\delta} \;\;\;\;\;\;
\delta ,\sigma ,m \;\;constans
 \]
It is easily shown that for $l=0$ we recover the Kramer
solution$^{34}$.  Now we make a transformation (23) with
\[ c=\left( \begin{array}{lll}
q&0&-s\\
0&1&0\\
-s&0&q\end{array}\right)
\]
and set $\delta =1/2.$  We get the Belinsky-Ruffini solution$^{35}$
which was obtained by inverse scattering method (see also ref.[15]).

\indent
c) Dobiasch-Maison and Cl\'{e}ment solutions

\indent
The formalism of Maison$^{12}$ is related with matrix (13) though a
transformation (23), with
\[ c=\left( \begin{array}{lll}
0&1&0\\
1&0&0\\
0&0&-1\end{array}\right)
\]
Dobiasch and Maison obtained a set of solutions with this formalism$^{36}.$
They studied the cases of Table IV for an Harmonic map given by
\[
\lambda =\frac{1}{a} ln\left(\frac{r-b-\frac{a}{2}}{r-b+\frac{a}{2}}\right)
\]
Using the same formalism Cl\'{e}ment applied the method of ``subspaces'' for
$\beta = 0$ and studied the cases of Table V and VI for some
$\lambda's^{27}.$

\indent
d) Chodos-Detweiler solution

\indent
Solutions obtained from subspaces in space time (Table III, V and VII) are less
know.  One example is the Chodos-Detweiler solution [37] that in our notation
(comparing it with (3)) is
\[
\gamma =\left(\begin{array}{lll}
e^\beta r^2 sin^2\theta &0&0\\
0& -e^\mu& A\\
0& A&\phi^2
\end{array}\right)
\]
\[ e^{\beta /2}=(1-{\beta^{2}\over{r^{2}}})/\psi \;\;\; , e^{\mu}=a_{2}
\psi^{p_1}+a_{1}\psi^{p_2},
\]
\[
A=(-a_{1}a_{2})^{1/2}(\psi^{p_1}-\psi^{p_2}), \;\;\;\;
\phi^{2}=a_{1}\psi^{p_1}+a_{2}\psi^{p_2} \;\; ,\;\;\;
\psi =\left({r-B\over{r+B}}\right)^{\lambda /2B}
\]
with
\begin{equation}
k=\frac{4}{\lambda^{2}}(4B^{2}-\lambda^{2}) \;\; , \;\;\;
p_{1,2}=1\pm \sqrt{1+k}
\end{equation}
The determinant of matrix $\lambda$ in (32) is
\[
-\alpha^{2} =- (1-{B^{2}\over{r^{2}}})r^{2}sin^{2}\theta
\]
which fulfills of course equation (5).  We writte now the corresponding
matrix g$=-\alpha^{-2/3}\gamma$
\begin{equation}
\hbox{g}=\left(\begin{array}{lll}
-\frac{\alpha^{\frac{4}{3}}}{\psi^2}&0&0\\
0&\frac{1}{\alpha^{\frac{2}{3}}}(a_2\psi^{p_1}+a_1\psi^{p_2})
&\frac{1}{\alpha^{\frac{2}{3}}}(-a_1a_2)^{\frac{1}{2}}(
-\psi^{p_1}+\psi^{p_2})\\
0&\frac{1}{\alpha^{\frac{2}{3}}}(-a_1a_2)^{\frac{1}{2}}(-\psi^{p_1}
+\psi^{p_2})&
\frac{1}{\alpha^{\frac{2}{3}}}(-a_1\psi^{p_1}-a_2\psi^{p_2})
\end{array}\right)
\end{equation}
\indent
We want to compare it with some element of Table II.  If we define
\[
b=a_{2}, c=a_{1}, \;\;\;{\gamma\over{q_{1}-\beta}}=-
\sqrt{-{a_{1}\over{a_{2}}}} \;\;\; , {\gamma\over{q_{2}-\beta}}=
\sqrt{-{a_{2}\over{a_{1}}}}
\]
\[
k={4(\alpha\beta -\delta\gamma)\over
     {(\alpha +\beta)}} \;\; , \;\; \tau ={1\over{\alpha+\beta}}{\it ln}
({r-B\over{r+B}})^{\lambda /B}\;\;\;and \;\;\;\;\lambda ={\it ln}\alpha
\]
we see that matrix (33) is just the matrix i) in Table II; so solution
(32) belongs to case i) in Table V.

\indent
e) Gross-Perry-Sorkin solution

\indent
The case q=0 of Table IV was studied in ref. [38].  We found that in this
case the metric can always be written as
\[
dS^{2}={1\over{I^{2}}}d\Lambda^{2}-dt^{2}+I^{2}(A_{3}d\phi + dx)^{2}
\;\;\; , \;\; I^{2}={4b^{2}\over{c-b\lambda}}
\]
\begin{equation}
d\Lambda^{2}=(1-\frac{2m}{r}+\frac{m^2sin^2\theta}{r^2})
(\frac{dr^2}{1-\frac{2m}{r}}+r^{2}d\theta^{2})+\left(1-\frac{2m}{r}\right)
r^2sin^2\theta \; d\phi^2
\end{equation}
in Boyer-Lindquist coordinates $\rho =\sqrt{r^{2}-2mr}sin\theta\;\;,\zeta =
(r-m)cos\theta$ .  Here we found again monopoles, dipoles, quadripoles etc.
solutions of 5D-gravity.  Let be $m=0$ in (34) and set
$b=-1/\sqrt{2}$,$c=2,\lambda =-16{\it l}b/r,$ what we find is just the
Gross-Perry-Sorkin magnetic monopole solution$^{39}$.  A magnetic monopole
solution with a Schwarzschild-like four dimensional metric is given in
ref. [40] which was generated from case b) in table IV and a very good
study of the geodesics of this metric was carryed out by Sarmah and
Krori$^{41}$.
\section{Conclusions.}
\indent
We have shown that the field equations of five dimensional gravity reduce to
harmonic maps eather for the space time as for the potential space
formalism.  We have supposed that the 5D-Riemannian space P contains three
Killing vectors, one of them being time-like (stationarity).  Nevertheless
this process can be done even for three space-like Killing-vectors, at less
in the space time, we should use then two coordinats $\xi = x +t$ and
$\eta =x-t$ in place of $z=\rho +i\zeta,$ and the metric (2) would be
\[
dS^2=H(x,t)(dx^{2}-dt^{2})+g_{ij}(x, t)dx^{i}dx^{j}
\]
The calculations would be the same.

\indent
If you wish an exact solution of 5D-gravity depending on two coordinates,
follow the following recipe:

\indent
1)Choose one or better two harmonic maps $\lambda_{1}$ and $\lambda_{2}$
in a flat space $V_{1}$ and a pair of harmonic maps $\lambda$ , $\tau$ in a
$V_{2}$ space (22).

\indent
2) Substitute $\lambda_{1}$ in Tables III and IV and obtain three solutions
in spacetime and six solutions in potential space.

\indent
3) Repeat 2) with $\lambda_{2}$

\indent
4) Substitute $\lambda_{1}$ and $\lambda_{2}$ in Tables V and VI and obtain
five solutions in space time and eicht in potential space.

\indent
5) Finally substitute $\lambda$ and $\tau$ in Table VII and obtain two
solutions in space time and two in potential space.

\indent
6) For calculating the function H of metric (2) substitute $\lambda_{1}$ or
$\lambda_{2}$ in (27) for the solutions of Table III, $\lambda_{1}$ and
$\lambda_{2}$ in (28) for the solutions of Table V and $\lambda ,\tau$ in (29)
or (30) for the space time solutions of Table VII for the groups
SL(2,{\bf R} ) or SO(2,1) respectively.

\indent
If you are not satisfied, make a (23)-transformation for each solution of
Tables IV, VI and VII you have.  You will obtain a new solution for each
matrix c for each seed solution you start from.

\indent
Some remarks about table VII must be done.  5D-gravity reduces to Einstein
theory in vacuum if we set $I=\kappa =1$ and the electromagnetic
potential vanishes.  This is so for the cases SL(2,{\bf R} ) in Table
VII, there the Einstein equations in vacuum are contained$^{19}.$
One of the isometry subgroups of Einstein-Maxwell theory
in the potential space is the group 0(2,1).  Electrostatic
and magnetostatic are contained here (and related by Bonnor
transformations$^{42}).$  Nevertheless if we only set $I=\kappa =1$, 5D
gravity reduces to Einstein-Maxwell theory with the restriction
\begin{equation}
f_{\mu\nu}f^{\mu\nu}=0
\end{equation}
where $f_{\mu\nu}$ is the Maxwell tensor.  In the Einstein-Maxwell theory
(35) means that we have fix the duality rotations.
\begin{equation}
F_{\mu\nu}=f_{\mu\nu}cos\alpha -f^{*}_{\mu\nu}sen\alpha
\end{equation}
$(f^{*}_{\mu\nu}={1\over{2}}\eta_{\mu\nu\alpha\beta}f^{\alpha\beta})$
for $\alpha =\pi /4.$ Then tensor $F_{\mu\nu}$ would fulfill restriction
(35).  Let $(g_{\mu\nu},f_{\mu\nu})$ be an Einstein-Maxwell static
solution, with Ernst potentials ${\cal E}=\bar{\cal E}$ and
$\phi ={\bar\phi}.$  Then using the subgroup $SO(2,1)$ in table VII, it can be
shown$^{28}$ that $\kappa = 1$,$\psi = a\phi$ ,$\chi =b\phi$ ,
$\epsilon =-c\phi^{2}$
and $f={\cal E}+\phi^{2}$ is an exact solution of this subspace in potential
space (see also ref. [29]) and
$(g_{\mu\nu} ,F_{\mu\nu}={1\over{\sqrt{2}}}(f_{\mu\nu}-f^{*}_{\mu\nu}))$
is an exact solution in
the space time.  Observe that these solutions in 5D-gravity are electrostatic-
magnetostatic, with the same electric and magnetic charge, i.e. dyons$^{18}.$
Some examples of these solutions was obtained from the Reisner-
Nordstr$\ddot o$m solution$^{28,43}$ and from the Bonnor dipole
solution$^{44}.$
\vfil\eject

\end{document}